\documentclass[12pt]{article}
\usepackage{geometry,amsmath,amssymb,epsfig}
\newcommand{\mathe}{\mathrm{e}}
\newcommand{\tmem}[1]{{\em #1\/}}
\newcommand{\tmop}[1]{\ensuremath{\operatorname{#1}}}
\newcommand{\tmtextbf}[1]{{\bfseries{#1}}}

\begin{document}
\title{Simulating the All-Order Strong Coupling Expansion I: Ising Model Demo}
\author{
Ulli Wolff\thanks{
e-mail: uwolff@physik.hu-berlin.de} \\
Institut f\"ur Physik, Humboldt Universit\"at\\ 
Newtonstr. 15 \\ 
12489 Berlin, Germany
}
\date{}
 
\maketitle
\begin{flushright} HU-EP-08/34 \end{flushright}
\begin{flushright} SFB/CCP-08-70 \end{flushright}
\thispagestyle{empty}

\begin{abstract}
  We investigate in some detail an alternative simulation strategy for lattice
  field theory based on the so-called worm algorithm introduced by Prokof'ev
  and Svistunov in 2001. It amounts to stochastically simulating the strong
  coupling expansion rather than the usual configuration sum. A detailed error
  analysis and an important generalization of the method are exemplified \
  here in the simple Ising model. It allows for estimates of the two point
  function where in spite of exponential decay the signal to noise ratio does
  not degrade at large separation. Critical slowing down is practically
  absent. In the outlook some thoughts on the general applicability of the
  method are offered.
\end{abstract}
\newpage

\section{Introduction}

Since about a decade {\cite{worm0}} there has been a development in the
condensed matter community where a novel approach to the simulation of
statistical models was pioneered. One replaces the original partition function
given by a sum over certain field configurations by its untruncated series
representation, usually the strong coupling or hopping parameter expansion.
For any finite but arbitrarily sized system this is regarded as an alternative
statistical system that is simulated by a novel Monte Carlo technique. \
Originally the method has emerged in quantum models that are `made classical'
by using the Trotter formula which leads to world-line type formalisms. In
{\cite{prokofev2001wacci}} the approach was used for classical spin systems in
two and three dimensions{\footnote{I would like to thank Urs Wenger for
drawing my attention to this paper.}}. In our opinion this idea is very useful
also for investigating systems of interest to lattice field theory and
elementary particle physics and it might be applicable in suitably generalized
form even to fermions and gauge models. Ultimately it could perhaps enable
steps toward the long sought generalization of {\cite{Rossi:1984cv}} beyond
infinite coupling{\footnote{Similar hopes have been recently expressed by S.
Chandrasekharan in his talk at Lattice 2008.}}.

In this publication that is planned to be the first of a series, we
concentrate our numerical experiments on the 2$D$ Ising model as a prototype
system. While the extension to abelian scalar bosonic theories in any $D$
seems straight-forward -- some cases can already be found in
{\cite{prokofev2001wacci}} -- other generalizations will be non-trivial. The
purpose of this paper, apart from some reformulation, \ is on the one hand to
introduce an important generalization of biasing the sampling of the series.
On the other hand a more detailed investigation of the behavior of some
observables of interest and their comparison with standard methods is made. It
should have become clear above that we are not discussing just a new Monte
Carlo algorithm but an exact reformulation where physical observables have
completely different estimators, variances and autocorrelations which deserve
study. Another study of the worm algorithm in the Ising model was recently
presented in {\cite{Deng:2007jq}}. There a detailed study of the Monte Carlo
dynamics is given refining results in {\cite{prokofev2001wacci}}. While we
confirm the practical absence of critical slowing down on our lattices, we
here do not attempt to determine dynamical critical exponents.

For a first closer look at the method we define
\begin{equation}
  Z (u, v) = \sum_{\{s\}} \mathe^{\beta \sum_{l = \langle x y \rangle} s (x) s
  (y)} s (u) s (v) . \label{Zs}
\end{equation}
Here $x, y, u, v$ denote sites of a cubic periodic lattice in arbitrary
dimension $D$. The outer sum is over Ising configurations $\{s (x) = \pm 1\}$,
in the exponent we sum over all links $l$. This is the numerator of the two
point function that may be written as
\begin{equation}
  G (x - y) = \left\langle s (x) s (y) \right\rangle = \frac{Z (x, y)}{Z (z,
  z)},
\end{equation}
with an arbitrary lattice site $z$. Generalizing {\cite{prokofev2001wacci}} we
consider a new partition function
\begin{equation}
  \mathcal{Z}= \sum_{u, v} \rho^{- 1} (v - u) Z (u, v) . \label{Zuv}
\end{equation}
The pairs of sites $u, v$ are now the `phase space' and $0 < \rho (x) <
\infty$ is a weight function demanded to possess the lattice periodicity.
Constant positive factors in $\rho$ will be irrelevant and we adopt the
normalization condition at the origin
\begin{equation}
  \rho (0) = 1.
\end{equation}
Now the two point function is given by
\begin{equation}
  G (z) = \rho (z) \frac{\left\langle \left\langle \delta_{v - u, z}
  \right\rangle \right\rangle}{\left\langle \left\langle \delta_{u, v}
  \right\rangle \right\rangle} \label{Gest},
\end{equation}
where the double bracket $\left\langle \left\langle \ldots \right\rangle
\right\rangle$ denotes averages with respect to (\ref{Zuv})
\begin{equation}
  \text{$\left\langle \left\langle A \right\rangle \right\rangle$} =
  \frac{1}{\mathcal{Z}} \sum_{u, v} A (u, v) \rho^{- 1} (v - u) Z (u, v)
\end{equation}
for observables $A (u, v)$. Note that any dependence of mean values on the
weight $\rho$ is canceled in (\ref{Gest}), but statistical errors will depend
on it.

To actually simulate (\ref{Zuv}) we need $Z (u, v)$ as weights which are not
available in closed form. Inserting (\ref{Zs}) and enlarging the phase space
to $\{u, v, s\}$ would lead to oscillating over-all weights and would not be
useful. Instead here the strong coupling expansion may be inserted. In a
somewhat sketchy notation, that will become more precise later, we put
\begin{equation}
  Z (u, v) = k \sum_{\gamma \in \Gamma (u, v)} (\tanh \beta)^{d (\gamma)} .
\end{equation}
Here the strong coupling expansion of the Ising model is written in powers of
$\tanh \beta$ and $k$ is an unimportant constant. The set of strong coupling
graphs contributing to (\ref{Zs}) is named $\Gamma (u, v)$ and for each
element $\gamma$ of the set $d (\gamma)$ denotes the degree in $\tanh \beta$
of this contribution. Of course $\Gamma (u, v)$ includes disconnected graphs
that may wind around the torus etc. If we insert this representation into
(\ref{Zuv}) we may consider $\{u, v, \gamma$\} as phase space points that may
now be sampled probabilistically to estimate $\left\langle \left\langle \ldots
\right\rangle \right\rangle$. The very important result of
{\cite{prokofev2001wacci}} is that it is almost trivial to design an ergodic
update from {\tmem{joint essentially local moves}} of \ $\{u, v, \gamma$\}
under which observables as those appearing in (\ref{Gest}) are even (almost)
free of critical slowing down. Local here means in particular that in each
step $\gamma$ is deformed only in a minimal way close $u$ or $v$. A detailed
description of the moves will follow. For the graphs of the usual partition
function ('vacuum' graphs) alone an ergodic update procedure would be more
complicated, it is easier in the enlarged setup. At the same time (\ref{Gest})
shows that the extended moves produce information on the two point function if
we record the occurring $v - u$ in a histogram.

It may seem surprising that local moves can (almost) eliminate critical
slowing down. But this prejudice is of course based on the idea of sampling
spin configurations that are long distance correlated. Here as one approaches
criticality, bigger strong coupling graphs will be important and apparently
the algorithm `knows' this and may even be guided in addition by choosing
$\rho$ appropriately.

There also is an important difference with regard to internal symmetries, the
spin flip Z(2) in the Ising case. Spin-flip odd observables cannot be
addressed anymore. The contributing graphs reflect the symmetry manifestly and
the field variables that transform under the symmetry have been summed over.
This should not be seen as a drawback: In the ideal situation of knowing the
exact solution this would be the same.

The plan of the paper is as follows. In 2. we derive the dimer form of the
Ising model that explicitly parametrizes all strong coupling graphs. In
addition the update for this system is defined. In 3. some numerical results
are collected and discussed. We end in 4. on conclusions and an outlook, where
we offer some speculations on the possibilities to generalize this
reformulation and its simulation in various directions.

\section{Ising model demonstration}

\subsection{Dimer formulation}

For a single Ising bond the trivial identity
\begin{equation}
  \mathe^{\beta s (x) s (y)} = \cosh \beta \sum_{k = 0, 1} [\tanh (\beta) s
  (x) s (y)]^k
\end{equation}
holds. Using it on each link \ with independent dimer or bond variables $\{k_l
= 0, 1\}$ we rewrite (\ref{Zs}) (up to the factor $2^{- N_x}$ inserted for
convenience) as
\begin{equation}
  Z (u, v) = (\cosh \beta)^{N_l} 2^{- N_x} \sum_{\{k, s\}} \prod_l [\tanh
  (\beta) \prod_{x \in \partial l} s (x)]^{k_l} s (u) s (v)
\end{equation}
with the number of links (sites) $N_l$ ($N_x$). The notation $l = \langle x y
\rangle$ has been replaced by the boundary set $\partial l =\{x, y\}$.

Next, the spins are summed over and leave behind the constraint
\begin{equation}
  2^{- N_x} \sum_{\{s\}} \prod_l [ \prod_{x \in \partial l} s (x)]^{k_l} s (u)
  s (v) = \Theta (k ; u, v) \in \{0, 1\}.
\end{equation}
The result $\Theta$ factorizes into local constraints with
\begin{equation}
  \theta (k ; y) = \left\{ \begin{array}{lll}
    1 & \tmop{if} & ( \sum_{l, \partial l \ni y} k_l) = \tmop{even}\\
    0 & \tmop{else} & 
  \end{array} \right.
\end{equation}
and their complements
\begin{equation}
  \overline{\theta} (k ; y) = 1 - \theta (k ; y)
\end{equation}
combining to
\begin{equation}
  \Theta (k ; u, v) = \left( \right. \prod_{y \notin \{u, v\}} \theta (k ; y)
  \left. \right) \times \left\{ \begin{array}{ll}
    \theta (k ; u) \hspace{1em} \tmop{if} \hspace{1em} u = v & \\
    \overline{\theta} (k ; u)) \overline{\theta} (k ; v) & \tmop{else} .
  \end{array} \right.
\end{equation}
In words: If $u = v$ holds, the number of dimers touching at any site has to
be even. For $u \not= v$, these two sites have to be surrounded by an odd
number of dimers with all other sites being even.

The previous steps coincide with those made in a duality transformation
{\cite{Wegner:1984qt}}. There, usually with no insertion (or $u = v)$, one
would solve the constraint by expressing $k_l$ by (the dual of) the discrete
`exterior derivative' {\cite{druhl1982afd}} of a $D - 2$ `form' on the dual
lattice, yielding self-duality for $D = 2$, the dual gauge theory for $D = 3$,
etc. For the algorithm {\cite{prokofev2001wacci}}, however, we stay with the
constrained $\{k_l \}$, in a sense intermediate between the original Ising
model and its dual.

The partition function of the enlarged ensemble now reads
\begin{equation}
  \mathcal{Z}= \sum_{u, v, \{k_l \}} \frac{\Theta (k ; u, v)}{\rho^{} (v - u)}
  \mathe^{- \mu \sum_l k_l}  \label{Zens}
\end{equation}
where the dimer sum together with the constraint uniquely label the graphs
$\gamma \in \Gamma (u, v)$ and the total dimer number equals $d (\gamma)$. The
coupling has been traded for the dimer chemical potential $\mu$,
\begin{equation}
  \tanh \beta = \mathe^{- \mu} .
\end{equation}
Mean values of observables $A (k ; u, v)$, which may now also depend on $k_l$,
are given by
\begin{equation}
  \left\langle \left\langle A \right\rangle \right\rangle =
  \frac{1}{\mathcal{Z}}  \sum_{u, v, \{k_l \}} A (k ; u, v) \frac{\Theta (k ;
  u, v)}{\rho^{} (v - u)} \mathe^{- \mu \sum_l k_l} . \label{primobs}
\end{equation}

Simple diagnostic observables for first Monte Carlo experiments are the
internal energy, or equivalently the average nearest neighbor correlation
\begin{equation}
  E = \frac{1}{D} \sum_{\nu} G ( \hat{\nu}) \label{Ei}
\end{equation}
and the susceptibility
\begin{equation}
  \chi = \sum_z G (z), \label{chi}
\end{equation}
where we sum over all directions $\nu$ and $\hat{\nu}$ are the corresponding
unit lattice vectors. The transcription of $\chi$ is obvious,
\begin{equation}
  \chi^{- 1} = \frac{\left\langle \left\langle \delta_{u, v} \right\rangle
  \right\rangle}{\left\langle \left\langle \rho (v - u) \right\rangle
  \right\rangle}  \label{chiob} .
\end{equation}
Thus, in particular for $\rho \equiv 1$, the fraction of coinciding $u = v$
has a simple interpretation and vanishes for $\beta \geqslant \beta_c$ in the
thermodynamic limit.

The corresponding translation for $E$ is
\begin{equation}
  E = \frac{1}{2 D} \frac{\left\langle \left\langle \rho (v - u) \delta_{|u -
  v|, 1} \right\rangle \right\rangle}{\left\langle \left\langle \delta_{u, v}
  \right\rangle \right\rangle} . \label{E1ob}
\end{equation}
Alternatively it may be related to the mean dimer occupancy in the subset $u =
v$ by differentiating $\ln Z (z, z)$ with respect to $\beta$
\begin{equation}
  E = \mathe^{- \mu} + 2 \sinh (\mu) \frac{\left\langle \left\langle
  \delta_{u, v}  \frac{1}{N_l} \sum_l k_l \right\rangle
  \right\rangle}{\left\langle \left\langle \delta_{u, v} \right\rangle
  \right\rangle} . \label{E2ob}
\end{equation}

\subsection{Prokof'ev-Svistunov worm algorithm}

With the algorithm{\footnote{Originally with $\rho \equiv 1$ only.}} of
{\cite{prokofev2001wacci}} we sample the ensemble (\ref{Zens}). The name of
the algorithm presumably derives from the fact that the constraint $\Theta (k
; u, v) \not= 0$ requires a line of active dimers ($k_l = 1$) connecting the
ends $u, v$ of the worm. However the connecting sequence of links is not
unique, the worm is `fuzzy' so to speak with fixed head and tail only. We
therefore prefer the name PS-algorithm honoring its inventors.

In any case the PS algorithm exploits the fact that we can base an ergodic
Monte Carlo algorithm for the ensemble (\ref{Zens}) on merely the following
two types of simple elementary steps that we apply in alternating order:
\begin{itemize}
  \item type I: we pick one of the $2 D$ nearest neighbors of $v$ with equal
  probability and call it $v'$ and the connecting link $l$. The proposed move
  $v \rightarrow v'$ with the simultaneous adjustment $k_l \rightarrow 1 -
  k_l$ is accepted with the Metropolis probability
  \begin{equation}
    p_{\tmop{acc}} = \min \left( 1, \frac{\rho (v - u)}{\rho (v' - u)}
    \mathe^{\mu (2 k_l - 1)} \right),
  \end{equation}
  otherwise the old configuration is maintained. The practical implementation
  is largely based on precomputed tables. It is not necessary to also move $u$
  due to translation invariance.
  
  \item type II: if we encounter a configuration $u, v, \{k_l \}$ with $u = v$
  we `kick' $u = v$ together to a randomly chosen other lattice site with
  unchanged $\{k_l \}$ with the probability $0 < p_{\tmop{kick}} < 1$. For $u
  \not= v$ we do nothing in this step, which is the dominant case.
\end{itemize}
Each such I-II step requires O(1) operations, independent of the lattice size.
We call this compound a micro-step. For a sequence of $2 N_l$ micro-steps
ergodicity can be shown. Any two configurations may be connected with non-zero
probability by the `worm dismantling' all active dimers of the first
configuration and then `re-building' the second one. The jumps allow to move
from one connected component to the next. As usual in ergodicity proofs, this
is in general of course a rather academic process! Following
{\cite{prokofev2001wacci}} we employ $p_{\tmop{kick}} = 1 / 2$. Some brief
numerical experiments with values 0.3 and 0.7 showed that this has hardly any
influence. Metropolis acceptance rates in steps I were reasonably far from the
extreme values 0 and 1 in all our simulations below, usually between 0.5 and
0.6. In {\cite{Deng:2007jq}} it has been observed that a correct and efficient
algorithm can even be based on steps of type I alone ($p_{\tmop{kick}} = 0$)
which we have however not tried here.

\section{Numerical demonstrations}

We group together $N_x$ micro-steps calling this an iteration, during which we
accumulate a sub-histogram of occurring separations $v - u.$ This may be
interpreted as measuring the whole set of 1-bit observables $\{\delta_{v - u,
z} \}$ -- one for each separation \ $z$ -- after each micro-step, mostly
implicit zeros, of course. Thus we 'always measure' and never give away any
information. Such an iteration has a computational complexity comparable to a
sweep in standard algorithms. We separately record the contributions to
primary observables from each such iteration. These records are treated as a
usual time series and an autocorrelation analysis as described in
{\cite{Wolff:2003sm}} is made.

Most physically interesting observables are of the derived type, nonlinear
functions, typically ratios, of primaries, see (\ref{Gest}). Details on their
error estimation, including the definition of integrated autocorrelation times
for such functions, can also be found in {\cite{Wolff:2003sm}}. We only repeat
here that in our definition $\sqrt{2 \tau_{\tmop{int}}}$ is the ratio between
the true and the `naive' error. Although this is wide-spread, there are
alternative definitions in the literature that may coincide with this one only
for large $\tau_{\tmop{int}}$, a case not relevant here.

The tests below are in $D = 2$ using lattices of size $T \times L$, but they
are expected to be representative for larger $D$, too.

\subsection{Simulations at $\beta_c$ with $\rho \equiv 1$}

Results for $\chi$ in the form (\ref{chiob}) and $E$ from both (\ref{E1ob})
and (\ref{E2ob}) are reported. For each lattice size we have performed $10^6$
iterations after equilibration{\footnote{Here and below we generously spend
10\% of the total statistics for equilibration.}}. The values are compatible
for instance with {\cite{Wolff:1989gz}} and our results with integrated
autocorrelation times are listed in table \ref{tab1}. We confirm the
practically complete absence of critical slowing down for the observables
under study.

\begin{table}[htbp]
  \centering
  \begin{tabular}{|c|c|c|c|c|c|c|}
    \hline
    $T = L$ & $L^{7 / 4} / \chi$ & $\tau_{\tmop{int}, \chi}$ & $E_{} (
    \ref{E1ob})$ & $\tau_{\tmop{int}, E ( \ref{E1ob})}$ & $E ( \ref{E2ob})$ &
    $\tau_{\tmop{int}, E ( \ref{E2ob})}$\\
    \hline
    16 & 0.9166(28) & 0.813(14) & 0.7263(10) & 0.567(5) & 0.7267(5) &
    1.94(4)\\
    \hline
    32 & 0.9153(33) & 0.815(14) & 0.7171(9) & 0.535(6) & 0.7171(3) & 2.13(4)\\
    \hline
    64 & 0.9117(38) & 0.860(15) & 0.7116(8) & 0.517(5) & 0.7120(2) & 2.41(5)\\
    \hline
    128 & 0.9196(44) & 0.928(17) & 0.7096(7) & 0.508(3) & 0.7097(1) &
    2.60(6)\\
    \hline
    256 & 0.9102(49) & 0.933(17) & 0.7089(7) & 0.504(3) & 0.7083(1) &
    2.81(7)\\
    \hline
  \end{tabular}
  \caption{Values and integrated autocorrelation times for simulations at the
  critical point $\beta = \ln (1 + \sqrt{2}) / 2$ and various lattice sizes.
  \label{tab1}}
\end{table}

It is instructive to compare the achieved precision with that of the single
cluster simulations in combination with standard spin estimators that were
reported in {\cite{Wolff:1989gz}}. The energy $E$ is a useful monitor for
simulations but non-universal and not really of physical interest. We hence
concentrate on $\chi$ which contains long distance contributions. We first
compare the growth of $\tau_{\tmop{int}, \chi}$ here with the one quoted in
{\cite{Wolff:1989gz}} for the single cluster algorithm (1C). While both
algorithms are very fast, the PS simulations seem to have even smaller and
less rising $\tau_{\tmop{int}, \chi}$. Does this automatically mean that PS is
preferable for measuring $\chi$? As we estimate in a completely different
ensemble also the variance can be quite different. In {\cite{Wolff:1989gz}}
errors are quoted for $\chi$ together with the number of single cluster steps
performed. Using the average cluster size (that actually equals $\chi$) it is
an easy exercise to predict the errors for a single cluster run of $10^6$
steps per spin, a CPU effort similar to the one in this study. Taking $L =
256$ as an example, the result is that the error on $\chi$ from a single
cluster simulation would be about 6 times smaller.

Due to the simplicity of the estimator (\ref{chiob}) we immediately know the
variance
\begin{equation}
  v_{\chi^{- 1}, \tmop{PS}} = \left\langle \left\langle (\delta_{u, v} -
  \chi^{- 1})^2 \right\rangle \right\rangle = \chi^{- 1} - \chi^{- 2} \approx
  \chi^{- 1} .
\end{equation}
Given the known final errors and the fact that we effectively measure after
each micro-step, we can also conclude the integrated autocorrelation time
$\overline{\tau}_{\tmop{int}, \chi}$ in time units of micro-steps. It ranges
from about $\overline{\tau}_{\tmop{int}, \chi} \approx 9$ for $L = 16$ to
$\overline{\tau}_{\tmop{int}, \chi} \approx 53$ for $L = 256$. This means that
an effectively independent estimate for $\chi^{- 1}$ arises after moving
dimers (modifying the strong coupling graphs) in only a small part of the
volume. The point is that we are discussing an {\tmem{integrated}}
autocorrelation time and not {\tmem{the slowest mode}}. Our observable couples
for instance only weekly to remote parts of the graph. Nevertheless, as
discussed above, taking into account also the variance, no miracle in terms of
efficiency has happened! In {\cite{Deng:2007jq}} quite similar observations
have been made for $\chi$. In their comparison with the Swendsen-Wang cluster
algorithm PS fares however somewhat better.

\begin{figure}[htbp]
  \centering
  \epsfig{file=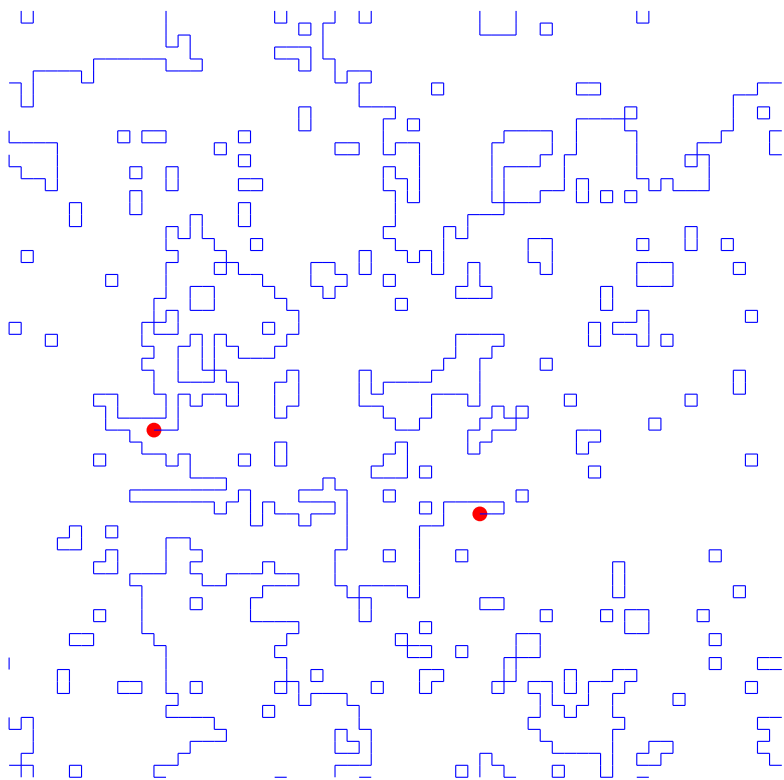,width=0.7\textwidth}
  \caption{A typical contribution for $T = L = 64, \beta = \beta_c$ without
  bias, $\rho \equiv 1$. The blobs are at $u$ and $v$, lines show $k_l = 1$.
  \label{config}}
\end{figure}

We end this subsection by visualizing a typical contributing graph in figure
\ref{config}.

\subsection{Biased PS: making use of $\rho$}

One use of the generalization of arbitrarily biasing the ensemble $(
\ref{Zens})$ with $\rho \not\equiv 1$ may be seen as follows. We assume to
be able to guess an approximate form of the two point function (up to a
factor) and use this for $\rho$. Then (\ref{Gest}) shows that the Monte Carlo
produces an approximately $z$-independent correction factor that turns our
guess into the exact correlation. This means, up to fluctuations and
imperfections of our guess, all our counting-bins of occurring $v - u$ will
have approximately equal filling. Apart from autocorrelation effects -- which
will be found to be minor -- this means that the {\tmem{relative}} statistical
error of correlations measured in this way will be $c / \sqrt{N}$ with the
number of iterations $N$ and essentially the same constant $c$ for all
separations. This implies a constant signal to noise ratio independent of the
size of the signal, a rather unusual situation.

We first tested this mechanism in the critical model simulated before. The
actual method to set $\rho$ was: In a first simulation with $\rho = 1$ where
we spent about 10\% of the planned total statistics, we have measured $G$
after some equilibration. Although not mandatory, we symmetrized the observed
histogram `by hand' over the geometric lattice symmetries: reflections along
the two directions and, if $T = L$ holds, cubic rotations. These symmetries
were, of course, only violated by the statistical fluctuations. With this
$\rho$ we then performed the main simulation{\footnote{In more difficult cases
one could think of a sequence of improving approximations.}}. We have indeed
observed almost flat $\left\langle \left\langle \delta_{v - u, z}
\right\rangle \right\rangle$ and compatibility with the known results.
Nevertheless there is no significant gain in precision which was not
unexpected: At criticality, in a finite volume, the two point function does
not decay exponentially but only by relatively moderate factors. The real
strength of the possibility of biasing toward larger separations is expected
in the massive disordered phase $\beta < \beta_c$ which we study next.

\subsection{Mass gap in the disordered phase}

In many applications one is interested in low-lying eigenvalues of the
transfer matrix. We pick a Euclidean time direction and resolve the lattice
site label into two-vectors $x = (x_0, x_1)$ labeling sites on a $T \times L$
torus and work in lattice units, $a = 1$ (integer $x_{\nu}$). We want to study
the temporal decay of correlations at fixed spatial momentum $p_1$
\begin{equation}
  G_{p_1} (x_0) = \sum_{x_1} G (x_0, x_1) \mathe^{- i p_1 x_1} .
\end{equation}
It is not difficult to see that in the dimer representation it is given by
\begin{equation}
  G_{p_1} (t) = \frac{\left\langle \left\langle \rho (t, v_1 - u_1)
  \delta_{v_0 - u_0, t} \mathe^{- i p_1 (v_1 - u_1)} \right\rangle
  \right\rangle}{\left\langle \left\langle \delta_{u, v} \right\rangle
  \right\rangle} .
\end{equation}
In the present study we restrict ourselves to observing the mass gap and probe
with $p_1 = 0$ only. Moreover, we found by a short test that for this
observable there is no disadvantage in letting $\rho$ only depend on time.
Then the correlation simplifies to
\begin{equation}
  G_0 (t) = \rho (t) \frac{\left\langle \left\langle \delta_{v_0 - u_0, t}
  \right\rangle \right\rangle}{\left\langle \left\langle \delta_{u, v}
  \right\rangle \right\rangle} .
\end{equation}
We now need a guess of the time-slice correlation and use this for $\rho$ and
have the Monte Carlo again only work out the corrections. It is sufficient to
histogram only the time separations and in addition count coincidences $u =
v$, if the normalization is needed. \ Due to symmetry we may combine $t$ and
$T - t$.

Below, instead of showing the correlation itself, we shall consider the time
dependent effective mass. Taking care of time periodicity we define $m (t + 1
/ 2)$ by numerically solving for $m$ in
\begin{equation}
  \frac{G_0 (t + 1)}{G_0 (t)} = \frac{\cosh (m (T / 2 - t - 1))}{\cosh (m (T /
  2 - t))}, \hspace{1em} m > 0, \hspace{1em} 0 \leqslant t < T / 2
\end{equation}
which for $m t \gg 1$ is expected to plateau at the asymptotic mass-gap of the
transfer matrix. This finite volume mass gap has a weak $L$-dependence and
approaches the true gap for large $m L$. Here we are not interested in this
second limit but just work around $m L \approx 5$ and below $\beta_c$.

\begin{figure}[htbp]
  \centering
  \epsfig{file=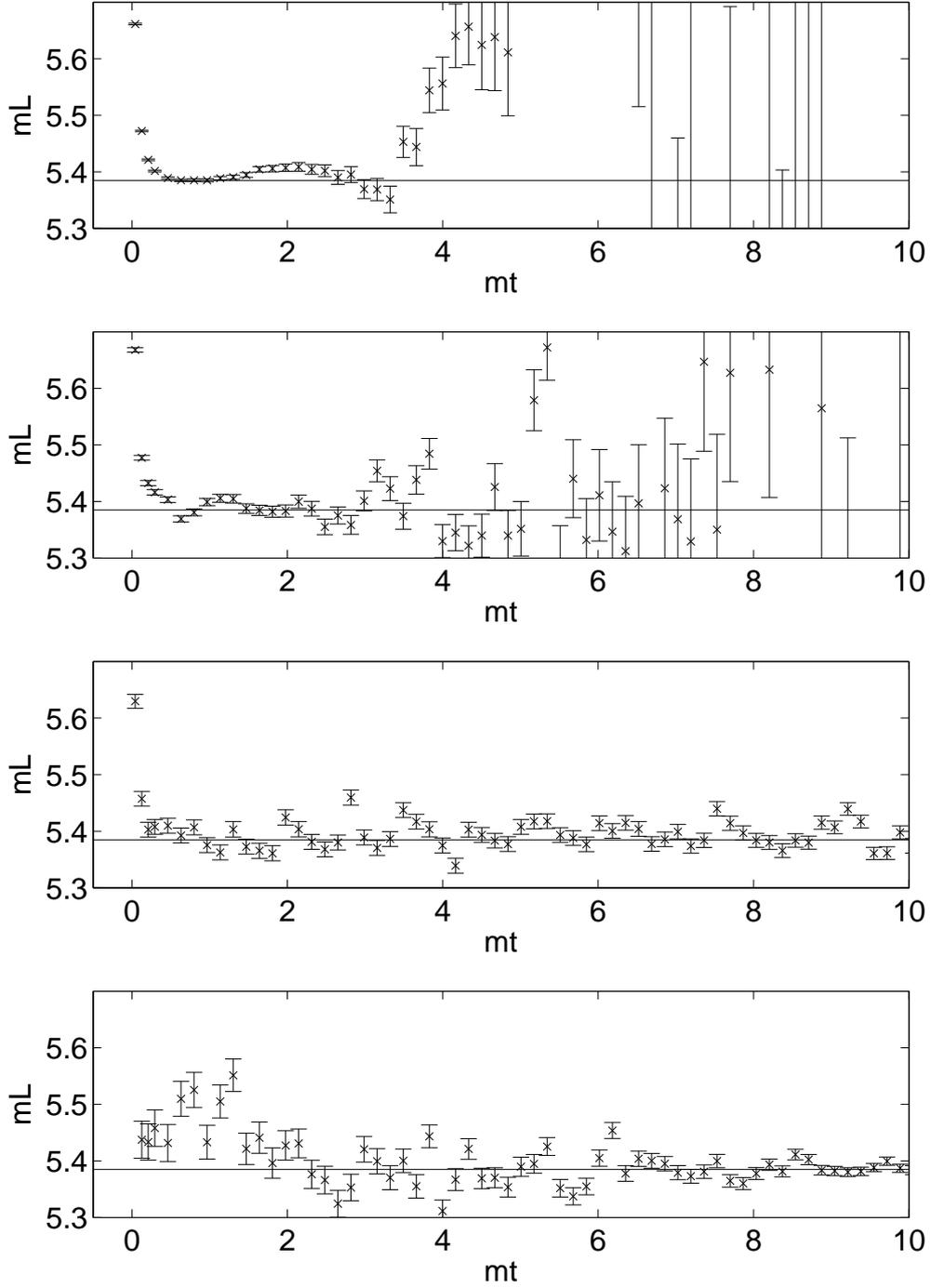,width=0.9\textwidth}
  \caption{Plots of the effective mass (times $L$) versus separation $t$ in
  correlation length units. From top to bottom: standard spin simulation,
  $\rho \equiv 1, \rho = \cosh (m (T / 2 - t))$, $\rho = \cosh ((m + 6 / T) (T
  / 2 - t))$. Errorbars are one sigma. \label{fig1}}
\end{figure}

For a test-bed with $L = 64, T = 256, \beta = 0.42$ we show in figure
\ref{fig1} the effective mass from several simulations. The horizontal lines
are the exactly known result{\footnote{I would like to thank Martin Hasenbusch
for advise with the Ising literature and for a C-code that evaluates the
finite volume mass-gap.}} from {\cite{Kaufman:1949ks}}
\begin{equation}
  m_{\tmop{ex}} [L = 64, \beta = 0.42] = 0.0841370 \ldots .
\end{equation}
Each of the plots corresponds to $10^6$ steps per lattice site. The uppermost
panel shows the standard situation in a favorable case. We have carried out
$10^6$ sweeps of the Swendsen-Wang algorithm and have measured the standard
spin-estimator. Due to our large statistics and a well isolated spectrum, we
see a clear plateau forming {\tmem{before}} the errors explode at larger
separation. As usual in such estimations, while the signal $G_0$ vanishes
exponentially the variance remains essentially constant. This could in
principle be improved by cluster estimators, but the signal to noise ratio
would still vanish exponentially with $t$, only at a slower rate.

In the next panel from above we see the result of a PS simulation with $\rho
\equiv 1$. The general impression is similar as before. The errors are
somewhat larger at small $t$ and smaller at large $t$. For the next simulation
we have biased for a flat histogram using an analytical guess $\rho \propto
\cosh (m (T / 2 - t))$ with a mass close to the expected one. We are rewarded
with a completely uniform error independent of $t$. In additional experiments
we tried to over-sample long distances and have used
\begin{equation}
  \rho \propto \cosh [(m + 2 n / T) (T / 2 - t)]
\end{equation}
which exponentially favors $v_0 - u_0 = T / 2$ over $v_0 = u_0$ by a factor
$\exp (n)$. The bottom plot in figure \ref{fig1} shows the case $n = 3$. We
indeed see even smaller errors at large $t$ (about a factor 0.6) where
systematic errors are smallest, a reversal of the standard situation where one
usually has to compromise between systematics and statistics. The three PS
runs took equal CPU time and the cluster simulation in our implementation was
a factor two longer, all on a single PC.

For extremely non-uniform choices of $\rho$ we expect to eventually run into a
problem, because $u, v$ have to travel over the whole lattice including all
separations for ergodicity. If one approaches such a case gradually, as we did
varying $n$, this should not set in abruptly but autocorrelation times should
start to rise. For the case above, and even for $n > 3$, we have hardly seen
such effects except that sometimes the summation window {\cite{Wolff:2003sm}}
for the autocorrelation function, in particular for $E$, had to be increased
to capture its tail. All observed integrated autocorrelation times of
effective masses were very close to $1 / 2$ in units of iterations. In
addition we notice that in the PS data the errorbars jump around the exact
line while the spin simulation results exhibit longer wavy structures. This
indicates that the PS data from different time separations are statistically
less dependent. Averaging the effective mass over a range or performing a fit
in some suitable window would probably reduce the error significantly. If we
take the deviations from the known constant for say $m t > 2$ and the
determined errors to form a $\chi^2$-value, we find reasonable values close to
the number of data points entering. We do however not attempt to optimize the
mass gap extraction any further here in this demonstration of principles.

\section{Conclusions and Outlook.}

We have investigated the worm algorithm of Prokof'ev and Svistunov in the
case of the $\tanh \beta$ expansion of the Ising model. We emphasize that this
involves a reformulation of the problem as a different statistical sum rather
than just a new simulation algorithm to produce thermalized spin configurations.
We confirm {\cite{prokofev2001wacci}} and {\cite{Deng:2007jq}} to the
effect that critical slowing down in practice is no issue for this method, at
least in the case studied here. The important generalization of biased
sampling ($\rho \not\equiv 1)$ allowed us to estimate the two point function
in the disordered massive phase with a signal to noise ratio that does not
decay with distance. Even the contrary can be achieved.

We expect that these features immediately generalize to other scalar models
with abelian field variables, e.g. $\phi^4$, Z($N$) clock models, U(1)
XY-model, Potts models in arbitrary dimension. A high precision study using
the PS algorithm for the phenomenologically important XY model in $D = 3$ was
already presented in {\cite{burovski:132502}}. In {\cite{endres:065012}} the
elimination of the sign-problem for bosons in a non-vanishing chemical
potential by the PS-method is demonstrated. For correlations beyond two point
functions one could consider more insertion points, perhaps with partially
restricted domains. For non-abelian spin models like O($N$), RP($N$), CP($N$),
$\tmop{SU} ( N ) \times \text{$\tmop{SU} ( N )$}$ the expansion and
constraints are clearly more involved but hopefully still manageable. It
should perhaps be kept in mind that for the continuum limit one is not
confined to the standard actions of these models. An immediate generalization
will be the world-line or loop-gas formulation of lattice fermions as
presented in {\cite{Wolff:2007ip}}. A similar representation can be written
for fermions in more than two dimensions. It will be all decisive for the
future usefulness of the approach whether the sign-problem can be handled
there. In two dimensions it is essentially absent and successful worm
simulations have in fact already been reported {\cite{Urs}}. This will allow
for the simulation of the O($N$) invariant Gross-Neveu model where along the
lines of {\cite{Wolff:2007ip}} we have encountered some problems with the
measurement of interesting correlation functions. Another important
generalization would be lattice gauge theory. There the insertion of spin
fields obviously should be generalized to Wilson loop insertions and gauge
invariance will be manifest term by term. In this way it will hopefully be
possible to efficiently generate the strong coupling graphs. The closed
surfaces of the vacuum graphs will be embedded in a larger class of graphs
with surfaces with boundaries as discussed to some degree in
{\cite{endres:065012}}. Also here the abelian case will be much simpler than
for example SU(2). First efforts towards a nonabelian dual formulation
(without so far employing a PS-type algorithm) can be found in
{\cite{Cherrington:2007is}}. The ultimate dream, of course, would be an
efficient alternative simulation technique for full QCD.\\
\tmtextbf{Acknowledgments:} I would like to thank Willi Rath and Urs Wenger
for discussions and Oliver B\"ar in addition also for useful comments on the
manuscript. Support by the Deutsche Forschungsgemeinschaft (DFG) in the
framework of SFB Transregio 9 is acknowledged.

\end{document}